\documentclass[a4paper,11pt]{article}
\pdfoutput=1

\usepackage{pos}

\usepackage{epsfig}
\usepackage{bbm}
\usepackage{bbold}
\usepackage{amsmath,amsthm}
\usepackage{yfonts}
\usepackage{mathrsfs}

\pdfoutput=1
\usepackage{color}
\usepackage{epsfig}
\usepackage{romannum}

\def\e{\varepsilon}

\newcommand{\wt}{\widetilde}

\usepackage{amsmath}
\usepackage{amsfonts}

\begin{document}

\def\a{\alpha}
\def\b{\beta}
\def\c{\chi}
\def\d{\delta}
\def\e{\epsilon}
\def\f{\phi}
\def\g{\gamma}
\def\h{\eta}
\def\i{\iota}
\def\j{\psi}
\def\k{\kappa}
\def\la{\lambda}
\def\m{\mu}
\def\n{\nu}
\def\o{\omega}
\def\p{\pi}
\def\q{\theta}
\def\r{\rho}
\def\s{\sigma}
\def\t{\tau}
\def\u{\upsilon}
\def\x{\xi}
\def\z{\zeta}
\def\D{\Delta}
\def\F{\Phi}
\def\G{\Gamma}
\def\J{\Psi}
\def\L{\Lambda}
\def\O{\Omega}
\def\P{\Pi}
\def\Q{\Theta}
\def\S{\Sigma}
\def\U{\Upsilon}
\def\X{\Xi}

\def\ve{\varepsilon}
\def\vf{\varphi}
\def\vr{\varrho}
\def\vs{\varsigma}
\def\vq{\vartheta}

\def\dg{\dagger}                                     
\def\ddg{\ddagger}                                   
\def\wt#1{\widetilde{#1}}                    
\def\mt{\widetilde{m}_1}
\def\mti{\widetilde{m}_i}
\def\rt{\widetilde{r}_1}
\def\mtt{\widetilde{m}_2}
\def\mttt{\widetilde{m}_3}
\def\rtt{\widetilde{r}_2}
\def\mb{\overline{m}}
\def\VEV#1{\left\langle #1\right\rangle}        
\def\be{\begin{equation}}
\def\ee{\end{equation}}
\def\ds{\displaystyle}
\def\ra{\rightarrow}

\def\bea{\begin{eqnarray}}
\def\eea{\end{eqnarray}}
\def\NO{\nonumber}
\def\Bar#1{\overline{#1}}


\def\pl#1#2#3{Phys.~Lett.~{\bf B {#1}} ({#2}) #3}
\def\np#1#2#3{Nucl.~Phys.~{\bf B {#1}} ({#2}) #3}
\def\prl#1#2#3{Phys.~Rev.~Lett.~{\bf #1} ({#2}) #3}
\def\pr#1#2#3{Phys.~Rev.~{\bf D {#1}} ({#2}) #3}
\def\zp#1#2#3{Z.~Phys.~{\bf C {#1}} ({#2}) #3}
\def\cqg#1#2#3{Class.~and Quantum Grav.~{\bf {#1}} ({#2}) #3}
\def\cmp#1#2#3{Commun.~Math.~Phys.~{\bf {#1}} ({#2}) #3}
\def\jmp#1#2#3{J.~Math.~Phys.~{\bf {#1}} ({#2}) #3}
\def\ap#1#2#3{Ann.~of Phys.~{\bf {#1}} ({#2}) #3}
\def\prep#1#2#3{Phys.~Rep.~{\bf {#1}C} ({#2}) #3}
\def\ptp#1#2#3{Progr.~Theor.~Phys.~{\bf {#1}} ({#2}) #3}
\def\ijmp#1#2#3{Int.~J.~Mod.~Phys.~{\bf A {#1}} ({#2}) #3}
\def\mpl#1#2#3{Mod.~Phys.~Lett.~{\bf A {#1}} ({#2}) #3}
\def\nc#1#2#3{Nuovo Cim.~{\bf {#1}} ({#2}) #3}
\def\ibid#1#2#3{{\it ibid.}~{\bf {#1}} ({#2}) #3}

\title{Testing an unstable cosmic neutrino background\footnote{Talk based on \cite{Dev:2023wel,preparation}. Slides can be found
at https://www.southampton.ac.uk/~pdb1d08/}}

\author*[a]{Pasquale Di Bari}

\affiliation[a]{School of Physics and Astronomy, University of Southampton\\
  Southampton, SO17 1BJ, U.K.}


\emailAdd{P.Di-Bari@soton.ac.uk}

\abstract{I discuss how different cosmological observations can test the possibility that neutrinos might be unstable on cosmological times, 
resulting into an unstable cosmic neutrino background. I also  discuss how actually there are different independent anomalies intriguingly hinting to 
such a possibility that would clearly point to new physics. 
I first focus on how the new DESI results place an upper bound on the sum of neutrino masses that starts to be in tension
with the lower bound from neutrino oscillation experiments and how this tension could be easily solved assuming unstable relic neutrinos. 
Then I show how 21 cm cosmology  allows to test radiative relic neutrino decays and how these could explain the controversial EDGES anomaly. 
I also discuss how the excess radio background and in particular the ARCADE 2 data can also be nicely explained by relic neutrino radiative decays. 
Finally, I point out the difficulties in building a model that does not clash with the upper limits on the effective magnetic moment coming from neutrino-electron scattering experiments and globular cluster stars.}

\FullConference{Proceedings of the Corfu Summer Institute 2024 "School and Workshops on Elementary Particle Physics and Gravity" (CORFU2024)\
12 - 26 May, and 25 August - 27 September, 2024\\
Corfu, Greece\\}


\maketitle

\section{Exploring the unknown}

Search of new physics is not just simply challenging but, with no evidence of new physics at colliders toward the end of the LHC Run III in 2026, 
it is becoming more and more clear that this will require in the next years not only formidable efforts but, likely, new ways 
to explore uncharted territories.  High luminosity LHC will only start in 2030 and 100 TeV colliders likely not before 2050.  
Moreover, new physics might well lie in a region of parameter space, at high energies and/or small couplings, 
well beyond  even the reach of future colliders.  In this situation, cosmological observations and neutrino physics provide alternative ways 
to access such remote regions. On the other hand, the remote regions we can explore are not within our direct control but are somehow scattered
in the energy versus couplings plane. It becomes then important to take advantage of all opportunities offered by experimental power
and nature itself. Fortunately, many new observational tools became  available in the last years or will soon become available. 
Not only low energy neutrino experiments
are making considerable progress but also cosmological and astrophysical observations provide many new avenues. First of all, the discovery
of gravitational waves has opened a new way to explore the physics of the early universe. In addition there are new observational tools that are 
providing new information with potential discoveries on the way: 21 cm cosmology experiments, CMB spectral distortions, the JWST telescope, new galaxy surveys
such as DESI. These new observations are not only placing more stringent constraints, but in many cases are hinting to new physics, beyond the standard models
of particle physics and cosmology.  In my talk I will focus on the constraints from the CMB spectrum, 21 cm cosmology, excess radio background. 

\section{Constraints from CMB spectrum}

The FIRAS instrument of COBE has measured the spectrum of the cosmic microwave background (CMB) in the frequency range (60 -- 600) GHz, corresponding approximately to the
photon energy range ($2.5 \times 10^{-5}$--$2.5\times 10^{-3}$)\,{\rm eV} and placed very strong upper bounds on deviations from a Planckian spectrum 
with temperature $T_{\gamma 0} = (2.7255 \pm 0.0006)\,{\rm K}$ \cite{Fixsen:2009ug}.  These translate into strong constraints in the parameter
of any model that predicts some amount of non-thermal radiation in that frequency range today. Interestingly, one can place
a lower bound on the lifetime of active neutrinos decaying radiatively into active neutrinos ($\nu_j \rightarrow \nu_i + \gamma $ with $i,j=1,2,3$). 
If the decaying active neutrinos are assumed to decay non-relativistically (for $m_i \gg T$), then the energy of the photon produced in the decay at the present time is simply given by
\begin{equation}
E_{\gamma 0} \simeq {m^2_j - m^2_i \over 2 m_j}\, {1 \over 1+ z_{\rm D}} \,  , 
\end{equation}
where $z_{\rm D}$ is the redshift at the decay time and if $m_i \simeq m_j$. Since neutrino oscillation experiments measure $m^2_j - m^2_i$, one can 
put constraints on the lifetime $\tau({\nu_j \rightarrow \nu_i + \gamma})$ versus the lightest neutrino mass $m_1$, assuming normal ordering for definiteness
and because current data disfavour inverted ordering. For example, consider the case $\nu_2 \rightarrow \nu_1 +\gamma$. In the hierarchical limit,
for $m_1 \lesssim \sqrt{m^2_2 - m^2_1} \sim 10\,{\rm meV}$, one obtains the highest value of $E_{\gamma 0}$, for $z_{\rm D} =0$, given by 
$E_{\gamma} \simeq 5\,{\rm meV}$, that is beyond the detected FIRAS range. However, for $z_D \gtrsim 2$, there are still non thermal photons that 
would be detected and this yields a constraint on the lifetime $\tau(\nu_2 \rightarrow \nu_1+\gamma) \gtrsim 10^{20}\,{\rm s}$.  In the quasi-degenerate
case, for $m_1 \gtrsim m_{\rm sol}$, one has $E_{\gamma 0} \simeq [(m_j-m_i)/2][1/(1+z_{\rm D})]$ and, in particular, one obtains
$E_{\gamma 0} \lesssim 2.5 \times 10^{-4}\,{\rm eV}$ for $m_1 \gtrsim 0.1\,{\rm meV}$,  below the FIRAS frequency range. 
In this case there would be no constraints but on the other hand the current cosmological upper bound on the sum of the neutrino masses, 
$\sum_i m_i < 0.12\,{\rm eV}$ (95\% C.L.) implies $m_1 \lesssim 0.03\,{\rm eV}$ (95\% C.L.), so that this limit would not be realised. 

Radiative neutrino decays $\tau({\nu_j \rightarrow \nu_i + \gamma})$ necessarily imply a non-vanishing effective neutrino magnetic moment $\mu^{ij}_{\rm eff}$ 
and this is true even if $\nu_i$ is a sterile neutrino. One has indeed the following general relation connecting radiative neutrino decay rate
and effective neutrino magnetic moment \cite{Mohapatra:1998rq,Xing:2011zza}
\be\label{gammatau}
\Gamma_{\nu_j \rightarrow \nu_i + \gamma} =  { \mu^{ij}_{\rm eff}\over 8\pi}\, \left({m^2_j - m^2_i \over m_j}\right)^3  \,  .
\ee
In this way a lower bound on $\tau(\nu_j \rightarrow \nu_i+\gamma)$ translates into an upper bound on $\mu^{ij}_{\rm eff}$.
For example, the lower bound  $\tau(\nu_2 \rightarrow \nu_1+\gamma) \gtrsim 10^{20}\,{\rm s}$ valid for $m_1 \lesssim m_{\rm sol} \simeq 
10\,{\rm meV}$, translates into an upper bound $\mu^{ij}_{\rm eff} \lesssim 5 \times 10^{-8}\mu_{\rm B}$, where 
$\mu_{\rm B} \equiv e \hbar/(2m_{\rm e})$ is the Bohr  magneton. For other active-to-active neutrino decay channels, one can obtain
slightly more stringent upper bound. However, these are in any case three-four orders of magnitude looser than the upper bounds
\be\label{upperbmu}
\mu^{ij}_{\rm eff}\lesssim 3.2 \times 10^{-11}\,\mu_{\rm B} \, , \;\;\; \mu^{ij}_{\rm eff}\lesssim 3 \times 10^{-12}\,\mu_{\rm B}
\ee
placed, respectively, by neutrino-electron scattering experiments \cite{Beda:2010hk} and globular 
cluster stars \cite{Raffelt:1990pj}. 

The Primordial Inflation Explorer (PIXIE) experiment will greatly improve FIRAS constraints on CMB spectral distortions 
and, correspondingly, the lower limits on lifetimes and the upper limit on magnetic moment placed by CMB will improve, respectively,
by four and two orders of magnitude. Moreover, the lower frequency threshold will decrease down to $\sim 30\,{\rm GHz}$.
In this way there is a consistency between the upper bound on $\mu^{ij}_{\rm eff}$ obtained from CMB spectral distortions and 
those obtained from neutrino-electron scattering experiments  and globular cluster stars.  

Notice that, conversely, from Eq.~(\ref{gammatau}) one can also translate the upper bound on the effective magnetic moment into a
very stringent lower bound on the lifetime for radiative neutrino decays  \cite{preparation}
\be\label{lbtau}
\tau_{\nu_i\rightarrow \nu_j +\gamma} \gtrsim 2.5 \times 10^{21}\,{\rm s}\,
\left({{\rm eV}\over \Delta m_{ij}} \right)^3 \,  .
\ee
For values $\Delta m_{ij} \ll 1 \,{\rm eV}$, one obtains such long lifetimes that it is difficult to think of any 
cosmological observational opportunity to test them. Therefore, the upper bounds on the effective magnetic moment,
represent a strong constraint on any cosmological application of relic neutrino radiative decays.  We will be back
on this important point in connection to an explanation of the excess radio background. 

\section{Cosmological tensions}

There are different tensions in current data from cosmological observations within the $\Lambda$CDM model.
The most famous tension is the Hubble tension \cite{DiValentino:2021izs}.  
The new JWST data seemed to solve the tension \cite{Freedman:2024eph} but it has been
noticed that, actually, they are not yet statistically significant so that the tension still persists \cite{Riess:2024vfa}.  There is not a clear simple  way to solve
the Hubble tension. Extensions of the $\Lambda$CDM model can ameliorate it but not solve it completely \cite{Schoneberg:2021qvd}. Systematic uncertainties
or local effects also do not seem enough to fully solve the tension. It might then be that this might be the result of a combination of different effects.
The new DESI results on baryon acoustic oscillation (BAO) also support some cosmological tensions, though of different nature. While BAO data 
are compatible with a value of $H_0$ consistent with the value inferred by {\em Planck} data from CMB anisotropies, they support a model 
of dark energy different from a simple cosmological constant, with a dependence of the dark energy equation of state parameter on redshift. 

However, there is another important tension emerging from DESI data that is relevant for our discussion. 
When these are combined with CMB data from {\em Planck} and ACT, the DESI collaboration obtains
the following upper bound on the sum of neutrino masses  within the $\Lambda$CDM model \cite{DESI:2024mwx} 
\begin{equation}\label{upperb}
\sum_i m_{\nu_i} \leq  72\,{\rm meV}  \;\;\; (95\% \, {\rm C.L.})     \,   .
\end{equation}
The best fit is obtained for $\sum_i m_{\nu_i} = 0$ when a prior $\sum_i m_{\nu_i} \geq 0$ is imposed.
This upper bound is in clear tension with the lower bound imposed by neutrino oscillation experiments \cite{Esteban:2024eli}
\be
\sum_i m_{\nu_i} \geq  58 \,{\rm meV} \,  .
\ee
This upper bound (\ref{upperb}) has been revisited taking into account new {\em Planck}  likelihoods 
resulting into a more relaxed one $\sum_i m_{\nu_i} < 120\,{\rm meV}$ at 95\% C.L. \cite{Allali:2024aiv}, implying a milder tension. 
However, even taking into account these different likelihoods for CMB data, 
new DESI results (DESI DR2 BAO + DR1 Full Shape) make the tension even stronger, 
since the upper bound on the sum of neutrino masses Eq.~(\ref{upperb}) gets even more stringent 
\cite{DESI:2025zgx,DESI:2025ejh}
\begin{equation}\label{upperb2}
\sum_i m_{\nu_i} \leq  64\,{\rm meV}  \;\;\; (95\% \, {\rm C.L.})    \,   .
\end{equation}
The same DESI data seem to suggest that an extension of the $\Lambda$CDM model with an evolving dark energy equation of state parameter
would solve the tension. Within a `$w_0w_a$CDM' model the upper bound gets indeed relaxed to
\begin{equation}\label{upperb3}
\sum_i m_{\nu_i} \leq  163\,{\rm meV}  \;\;\; (95\% \, {\rm C.L.})     \,   .
\end{equation}
However, when SN data are also combined the bound gets more stringent again and the DESI collaboration obtains (suing Pantheon+ data for SN)
\begin{equation}\label{upperb4}
\sum_i m_{\nu_i} \leq  117\,{\rm meV}  \;\;\; (95\% \, {\rm C.L.})     \,   ,
\end{equation}
so that some tension still persists considering that the best fit is still found for vanishing neutrino masses.
A  possible  solution of this tension is to consider unstable relic neutrinos. 
The cosmological upper bound on the sum of neutrino masses we reported assumes  neutrino lifetimes
$\tau_{\nu_i} \gg 0.1\,t_0$, where $t_0 \simeq 4 \times 10^{17}\,{\rm s}$ is the age of the universe.  If all ordinary neutrinos 
decay with shorter lifetimes, then their role as hot dark matter can be neglected and the upper bound gets strongly relaxed \cite{Escudero:2020ped,Craig:2024tky}.
The neutrino lifetimes should anyway be longer than $\sim 10^6 \, (\sum_i m_{\nu_i}/50\,{\rm meV})^5$ 
not to have a clash with CMB anisotropies observations that support the presence of neutrino free streaming \cite{Barenboim:2020vrr}.
Moreover, from the CMB spectral distortion constraints we discussed, such short lifetimes necessarily imply that neutrinos have to decay invisibly. 
For example, if active neutrinos interact with a scalar field $\phi$ with interactions \cite{Escudero:2020ped,Craig:2024tky}
\be
{\cal L}_{\nu-\phi} = {\lambda_{ij} \over 2}\,\bar{\nu}_i \nu_j \phi  +{\rm h.c.} \,   ,
\ee 
one would open the decay channel $\nu_i \rightarrow \nu_j + \phi$ with lifetime
\be
\tau_{\nu_i\rightarrow\nu_j +\phi} \simeq 7 \times 10^{17}\,{\rm s} \left({0.05\,{\rm eV} \over m_{\nu_i}}\right)
 \, \left({10^{-15} \over \lambda_{ij}^2}\right)^2 \,   .
\ee
Therefore, the tension between oscillation neutrino experiments and cosmological observations 
might be interpreted as if it suggests the existence of a low scale dark sector destabilising the 
cosmic neutrino background.

\section{Radiative neutrino decays: specific intensity}

We have seen that FIRAS constraints place stringent lower bound on the lifetime of active-to-active radiative neutrino decays since
the photon energy necessarily lies within the FIRAS range. However, if the active neutrino $\nu_i$ decays into a sterile neutrino $\nu_0$, 
then neutrino oscillation experiments do not constraint $m^2_i - m^2_0$ and the photon energy can be below the FIRAS low threshold 
energy $E_\gamma^{\rm FIRAS, low} = 60\,{\rm GHz} = 2.5 \times 10^{-4}\,{\rm eV}$.  In this case, as we are going to discuss in next sections, one can
consider different constraints coming from 21 cm cosmological global signal and excess radio background.

A quantity that describes non-thermal radiation  is the {\em specific intensity}
\be\label{specificint}
I_{\g_{\rm nth}}(E,z) \equiv {d{\cal F}^{\g_{\rm nth}}_E \over dA\,dt \, dE \,d\Omega}  = {1 \over 4\pi} \, {d\varepsilon_{\gamma_{\rm nth}} \over d E}
= {E^3 \over 4\,\pi^3}\,[e^{E/T_{\g_{\rm nth}}(E,z)}-1]^{-1} \,  .
\ee
In the last expression $T_{\g_{\rm nth}}(E,z)$ is the {\em effective (radiometric) temperature} of non-thermal radiation, corresponding to the
temperature of a thermal (black body) radiation with the same specific intensity at frequency $\nu = E/(2\pi)$. 
For $E \ll T_{\g_{\rm nth}}$, one obtains a simple linear relation between effective temperature and specific intensity
\be
T_{\g_{\rm nth}} \simeq {4\pi^3 \over E^2} \, I_{\g_{\rm nth}}(E,z) \,  .
\ee
In the case of radiative neutrino decay one obtains for the specific intensity \cite{Masso:1999wj,Chianese:2018luo,Dev:2023wel}:
\be\label{Inth}
I_{\gamma_{\rm nth}}(E,z)  = {1 \over 4 \pi} \, {d\varepsilon_{\g_{\rm nth}} \over d E}
\, = {n_{\nu_i}^{\infty}(z) \over 4\,\pi} 
{e^{-{t(a_{\rm D}) \over \tau_i}} \over H(a_{\rm D}) \, \tau_i} \,  ,
\ee
where $\tau_i$ is the lifetime of the decaying neutrinos and $\varepsilon_{\g_{\rm nth}}$ is the energy density of the non-thermal radiation. The quantities $H(a_{\rm D})$ and $t(a_{\rm D})$ are, respectively, the expansion rate and the age of the universe calculated at the time of decay of the relic neutrinos that produced photons with energy $E$ and detected at redshift $z$. The time of decay $t_{\rm D} \equiv t(a_{\rm D})$
corresponds to a redshift $z_{\rm D} = a_{\rm D}^{-1} -1$  
and scale factor $a_{\rm D} = (E  /\Delta m_{0i})\, a \leq  a$. Finally,
\be\label{nnu}
n^{\infty}_{\nu_i}(z) = {6\over 11}{\zeta(3)\over \pi^2}\,T^3(z)
\ee
is the relic neutrino number density calculated at the time of detection, at redshift $z$, 
in the standard stable neutrino case, where $T$ is the standard photon temperature.

The expansion rate at the decay, $H(a_{\rm D})$, can be calculated in the $\L$CDM model  as
\be
H(a_{\rm D}) = H_0\,\sqrt{\Omega_{{\rm M}0} \,a_{\rm D}^{-3} + \Omega_{\Lambda  0}} = 
H_0 \, \sqrt{\Omega_{M0}} \, a_{\rm D}^{-{3\over 2}} \, 
\left(1 + {a_{\rm D}^3 \over a_{\rm eq}^3} \right)^{{1 \over 2}}  \,  ,
\ee
where $a_{\rm eq} \equiv (\Omega_{M0}/\Omega_{\Lambda 0})^{1/3} \simeq 0.77$, 
$\Omega_{M0} \simeq 0.3111$, $H_0 \simeq t_0^{-1}$ and $t_0 \simeq 13.8 \, {\rm Gyr} \simeq 4.35 \times 10^{17}\,{\rm s}$ \cite{Planck:2018vyg}. One can also obtain
an analytical expression for the age of the universe at the time of decay, $t(a_{\rm D})$ \cite{DiBari:2018vba}:
\be
t(a_{\rm D}) = {2\over 3}\,{H_0^{-1}\over \sqrt{\Omega_{\Lambda 0}}}\,
\ln\left[\sqrt{\left({a_{\rm D}\over a_{\rm eq}}\right)^3} + \sqrt{1+\left({a_{\rm D}\over a_{\rm eq}}\right)^3} \right] \,  .
\ee

\section{21 cm cosmology}

The 21 cm cosmological global signal represents an important diagnostic tool to test new physics after the recombination era \cite{Pritchard:2011xb}.
The 21 cm (emission or absorption) line is produced by hyperfine transitions between the 
spin-singlet and triplet energy levels of the 1s ground state of Hydrogen atoms. 
The energy splitting between the two levels is $E_{21} = 5.87\, \mu{\rm eV}$, corresponding 
to a rest frequency $\nu_{21}^{\rm rest}= 1.420\, {\rm GHz}$. 
This transition can be used cosmologically to obtain information on the cosmological history and parameters in a
wide redshift range $z \sim 7$--$200$. 
The 21 cm brightness temperature parameterises the brightness contrast 
between the cosmic radiation and the absorbed or emitted radiation in the 21 cm transitions 
is usually referred to as the 21 cm cosmological global signal and is approximately given by \cite{Zaldarriaga:2003du}
\be\label{T21}
T_{21}(z) \simeq  23\,{\rm mK} \, (1+\d_{\rm B})\, x_{H_I}(z) \,\left({\Omega_{{\rm B}0} h^2 \over 0.02}\right)\,
\left[\left({0.15 \over \Omega_{{\rm M}0}h^2}\right)\,
\left({1+z \over 10} \right)  \right]^{1/2}  \,\left[
1 - {T_\g(z) \over T_{\rm S}(z)} \right]  \,  ,
\ee
where $T_{\rm S}(z)$ is the spin temperature describing the  triplet-to-singlet state density ratio. 
In this expression $\d_{\rm B} = (\rho_{\rm B}-\bar{\rho}_{\rm B})/\bar{\rho}_{\rm B}$ is the fractional
baryon overdensity, $x_{H_I}(z)$ is the neutral hydrogen fraction.
If the spin temperature is equal to the photon temperature ($T_{\rm S} = T_{\gamma}$), 
then photons are absorbed and reemitted with the same intensity and there is no visible signal. 
Also, if all atoms are ionised so that $x_{H_I} = 0$, there cannot be any signal. 

The most prominent feature that is expected within the $\Lambda$CDM model in the 21 cm cosmological signal is an absorption 
signal, corresponding to a negative value of  $T_{21}$, at redshifts in the range $z=10$--$30$. The EDGES collaboration  
claimed to have detected such absorption feature  centred at $z = z_{\rm E} \simeq 17$ \cite{EDGES}, 
thus within the expected range of redshifts. However, the measured (negative) value of $T_{21}$ is approximately
twice the expected one. This anomalous signal can be interpreted in terms of a non-thermal radiation component produced by relic neutrino decays with 
a temperature, at $z =z_E$, $T_{\gamma_{\rm nth}} \simeq 60\,{\rm K}$. 
We can assume, for definiteness, that the decaying neutrinos are the lightest active neutrinos  decaying 
non-relativistically into quasi-degenerate sterile neutrinos. 
The Eq.~(\ref{Inth}) has to be specialised to the case $z = z_{\rm E}$ and $E = E_{21} = 5.87\mu {\rm eV}$ and imposing
$T_{\gamma_{\rm nth}}(z=z_E) \simeq 60\,{\rm K}$, one obtains 
\be\label{DmtauEDGES}
(\D m_1^{3/2}\tau_1)^{\rm EDGES} \simeq 4.0 \times 10^{13}\,\,{\rm eV}^{3/2}\,{\rm s}  \,  .
\ee
The EDGES anomaly is controversial and a few studies have suggested that the signal is contaminated by some foreground contribution,
for example originating in the ionosphere.  The SARAS3 experiment has even rebutted the EDGES claim \cite{Singh:2021mxo}, so we need to wait for 
more results. In this respect it is exciting that the moon-based experiment LuSEE will soon be able to measure the 21 cm cosmological global signal in absence of
Earth foregrounds  \cite{2023arXiv230110345B}. 

\section{Excess radio background}

The ARCADE 2 balloon-borne experiment has measured the absolute temperature of sky at frequencies in the range (3--10)\,GHz \cite{Fixsen:2009xn},
well below the FIRAS low threshold, and covering a $8.4\%$ portion of the sky.  Only 6 data points gave in the end 
a meaningful result not dominated by noise. These point clearly show an excess compared to the CMB  temperature that is not
described by a thermal component (that implies that the temperature changes with the frequency).  The excess cannot be
explained in terms of a known population of radio sources and different attempts have not detected any anisotropy so that the source of the
excess has to be extremely smooth and this seems to suggest that can be better described by a smooth background rather than unknown radio sources. 
It also disfavours solutions where the signal is somehow correlated with the dark matter distribution, since one would expect some level of anisotropy
that is excluded by the observations. It is quite fair to conclude that the `nature of the background is still unknown' \cite{Grebenev:2024nkw}, 
and for this reason the excess radio background should be regarded as a mystery. 

An intriguing possibility is that the source of non-thermal radiation is given by 
non-relativistic radiative relic neutrino decays into quasi-degenerate sterile neutrinos, as we discussed in the case of the 21 cm global signal.  
As for the 21 cm cosmological signal, we can still start from the general expression Eq.~(\ref{specificint}) and specialise it at the 
present time and this time, since we have different energies to fit, leave indicated the energy dependence (also contained in $a_{\rm D}$), finding
\be\label{Tgammanth}
T_{\gamma_{\rm nth}}(E,0) \simeq  {6\,\zeta(3)\over 11 \, \sqrt{\Omega_{{\rm M}0}}}\,
{T_{0}^3 \over E^{1 / 2}\, \Delta m_1^{3 / 2} } \, {t_0 \over \tau_1} 
\,\left(1 +{a_{\rm D}^3 \over a_{\rm eq}^3} \right)^{-{1 \over 2}} \,  .
\ee
When this is used to fit the six data points found by ARCADE 2, one finds as a best fit  \cite{Dev:2023wel}
\be
\tau_1 = 1.46  \times 10^{21}\,{\rm s}  \,  , \;\;\;\;  m_1 -m_{\rm s} = 4.0 \times 10^{-5}\,{\rm eV}
\ee
with a very good $\chi_{\rm min}^2 \simeq 1$.  The curve for the effective temperature for these best fit parameters  is shown in the figure 
with a orange thick solid line.  In the vertical axis $T_{\rm ERB}(E) \equiv T_{\gamma 0}(E) - T_{{\rm CMB},0}$ is the effective 
temperature of the excess radio background. It can be noticed how one of the most clear features of the fit is the existence of an
end-point at $E = m_1 -m_{\rm s}$.
\begin{figure}
\begin{center}
  \includegraphics[scale=0.75]{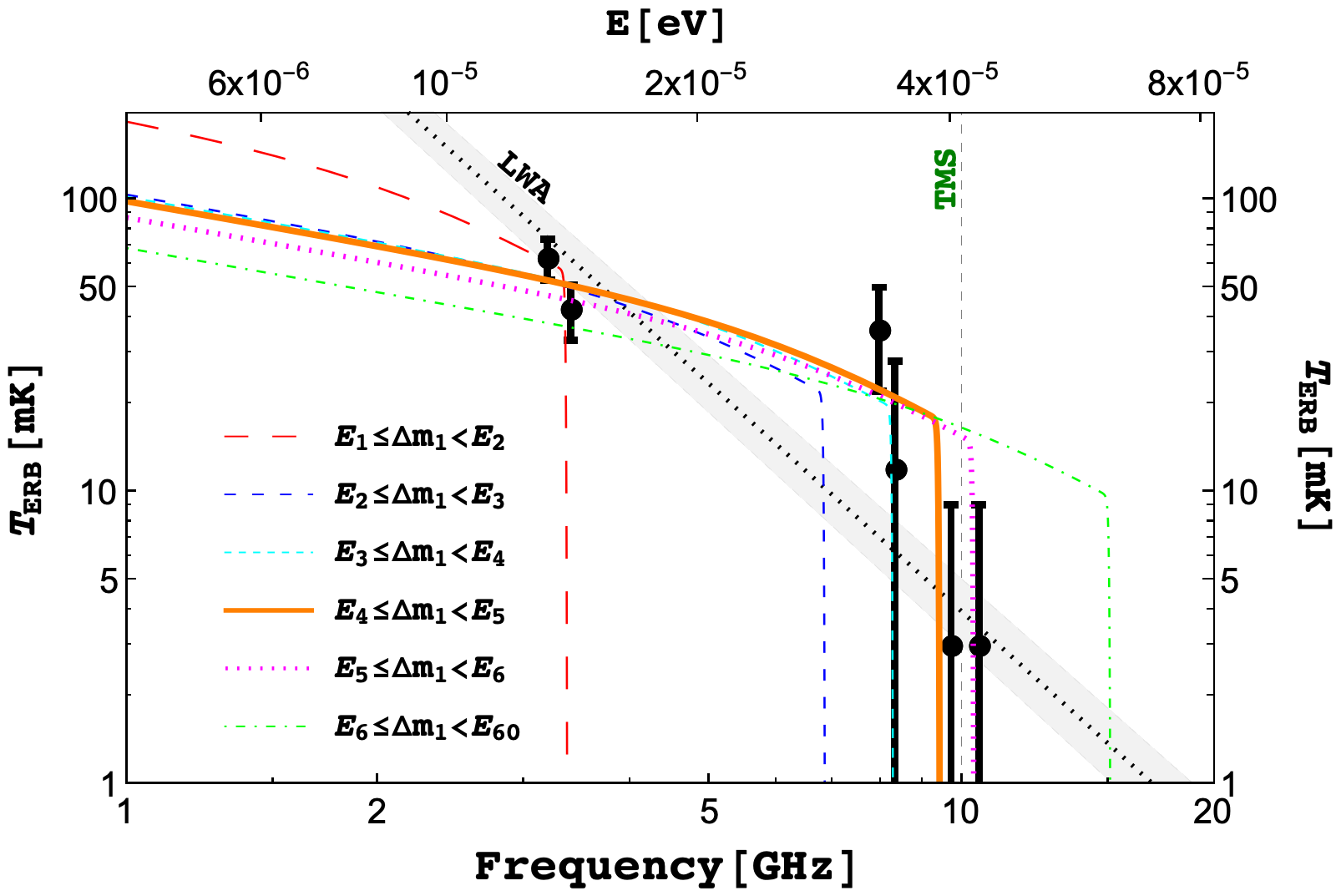}
\end{center}
    \caption{Best fit curves for $T_{\rm ERB}$ obtained with Eq.~(\ref{Tgammanth}). The thick solid orange curve corresponds to a solution very close to the
    best global fit ($\D m_1 = 4.0 \times 10^{-5}\,{\rm eV}$  and $ \tau_1= 1.46 \times 10^{21}\,{\rm s}$).
    The ARCADE 2 data points are taken from Ref.~\cite{Fixsen:2009xn}, while the power-law fit 
    $\beta = -2.58 \pm 0.05$ (dotted line with grey shade) is from \cite{Dowell:2018mdb}. The figure is taken from \cite{Dev:2023wel}.}
    \label{fig:bestfit}
\end{figure}
These best fit values correspond to
$\D m_1^{3/2}\tau_1 \simeq 4.0 \times 10^{14}\,\,{\rm eV}^{3/2}\,{\rm s}$, a value about 1 order of magnitude higher than the best fit value Eq.~(\ref{DmtauEDGES}) explaining the EDGES anomaly. Even taking into account the errors, there is a $\sim 3 \sigma$ tension. However, this is not a problem  since as we said the EDGES anomaly might be dominated by foregrounds and the anomalous signal one can predicts from the solution to the excess radio background
is much weaker than the one observed by EDGES. Of course, the opposite would have been a problem, since EDGES should have seen such a strong signal.

This result sounds very exciting but there is a clear challenge: it violates the lower limit on the lifetime in Eq.~(\ref{lbtau})
that in our case gives $\tau_i \gtrsim 10^{36}\,{\rm s}$.  A model should then be able to circumvent the connection between lifetime and
effective magnetic moment enforced by the relation (\ref{gammatau}).

\section{Final remarks}

\begin{itemize}
\item New exciting  cosmological tools allow to explore new physics in regimes, both energy and coupling-wise, inaccessible to colliders;
\item At very low scales there are interesting mysteries  that might be explained by an unstable relic neutrino 
background decaying invisibly and radiatively; 
\item The short requested lifetimes to solve the excess radio background, and possible the EDGES anomaly, are challenging to explain
without violating the upper bounds on the effective magnetic moment but this maybe makes things even more exciting;
\item We will have soon experiments testing these ideas, both lunar-based radio antennas that will try to detect the 
21 cm cosmological global signal and the new Tenerife microwave spectrometer (TMS) that will try to measure the excess radio
background at higher frequencies than ARCADE 2.
\end{itemize}

\vspace{-1mm}
\subsection*{Acknowledgments}

I acknowledge financial support from the STFC Consolidated Grant ST/T000775/1 and
from the European Union’s Horizon 2020 Europe research and innovation programme under  
the Marie Sk\l odowska-Curie grant agreement HIDDeN European  ITN project (H2020-MSCA-ITN2019//860881-HIDDeN).

I wish to thank Marco Chianese, Bhupal Dev, Rishav Roshan, Rome Samanta,  Ivan Martinez-Soler, for a very fruitful  collaboration.


\begin{thebibliography}{100}

\bibitem{Dev:2023wel}
P.~S.~B.~Dev, P.~Di Bari, I.~Mart\'\i{}nez-Soler and R.~Roshan,
{\em Relic neutrino decay solution to the excess radio background},
JCAP \textbf{04} (2024), 046
[arXiv:2312.03082 [hep-ph]].

\bibitem{preparation}
P.~S.~B.~Dev, P.~Di Bari, I.~Mart\'\i{}nez-Soler and R.~Roshan, in preparation.

\bibitem{Fixsen:2009ug}
D.~J.~Fixsen,
{\em The Temperature of the Cosmic Microwave Background},
Astrophys. J. \textbf{707} (2009), 916-920
[arXiv:0911.1955 [astro-ph.CO]].

\bibitem{Mohapatra:1998rq}
R.~N.~Mohapatra and P.~B.~Pal,
{\em Massive neutrinos in physics and astrophysics. Second edition},
World Sci. Lect. Notes Phys. \textbf{60} (1998), 1-397

\bibitem{Xing:2011zza}
Z.~z.~Xing and S.~Zhou,
{\em Neutrinos in particle physics, astronomy and cosmology}.

\bibitem{Beda:2010hk}
A.~G.~Beda, V.~B.~Brudanin, V.~G.~Egorov, D.~V.~Medvedev, V.~S.~Pogosov, M.~V.~Shirchenko and A.~S.~Starostin,
{\em Upper limit on the neutrino magnetic moment from three years of data from the GEMMA spectrometer},
[arXiv:1005.2736 [hep-ex]].

\bibitem{Raffelt:1990pj}
G.~G.~Raffelt,
{\em New bound on neutrino dipole moments from globular cluster stars},
Phys. Rev. Lett. \textbf{64} (1990), 2856-2858

\bibitem{DiValentino:2021izs}
E.~Di Valentino, O.~Mena, S.~Pan, L.~Visinelli, W.~Yang, A.~Melchiorri, D.~F.~Mota, A.~G.~Riess and J.~Silk,
{\em In the realm of the Hubble tension\textemdash{}a review of solutions},
Class. Quant. Grav. \textbf{38} (2021) no.15, 153001
[arXiv:2103.01183 [astro-ph.CO]].

\bibitem{Freedman:2024eph}
W.~L.~Freedman, B.~F.~Madore, I.~S.~Jang, T.~J.~Hoyt, A.~J.~Lee and K.~A.~Owens,
{\em Status Report on the Chicago-Carnegie Hubble Program (CCHP): Measurement of the Hubble Constant Using the Hubble and James Webb Space Telescopes},
[arXiv:2408.06153 [astro-ph.CO]].

\bibitem{Riess:2024vfa}
A.~G.~Riess, D.~Scolnic, G.~S.~Anand, L.~Breuval, S.~Casertano, L.~M.~Macri, S.~Li, W.~Yuan, C.~D.~Huang and S.~Jha, \textit{et al.}
{\em JWST Validates HST Distance Measurements: Selection of Supernova Subsample Explains Differences in JWST Estimates of Local H $_{0}$},
Astrophys. J. \textbf{977} (2024) no.1, 120
[arXiv:2408.11770 [astro-ph.CO]].

\bibitem{Schoneberg:2021qvd}
N.~Sch\"oneberg, G.~Franco Abell\'an, A.~P\'erez S\'anchez, S.~J.~Witte, V.~Poulin and J.~Lesgourgues,
{\em The H0 Olympics: A fair ranking of proposed models},
Phys. Rept. \textbf{984} (2022), 1-55
[arXiv:2107.10291 [astro-ph.CO]].

\bibitem{DESI:2024mwx}
A.~G.~Adame \textit{et al.} [DESI],
{\em DESI 2024 VI: cosmological constraints from the measurements of baryon acoustic oscillations},
JCAP \textbf{02} (2025), 021
[arXiv:2404.03002 [astro-ph.CO]].

\bibitem{Esteban:2024eli}
I.~Esteban, M.~C.~Gonzalez-Garcia, M.~Maltoni, I.~Martinez-Soler, J.~P.~Pinheiro and T.~Schwetz,
{\em NuFit-6.0: updated global analysis of three-flavor neutrino oscillations},
JHEP \textbf{12} (2024), 216
[arXiv:2410.05380 [hep-ph]].

\bibitem{Allali:2024aiv}
I.~J.~Allali and A.~Notari,
{\em Neutrino mass bounds from DESI 2024 are relaxed by Planck PR4 and cosmological supernovae},
JCAP \textbf{12} (2024), 020
[arXiv:2406.14554 [astro-ph.CO]].

\bibitem{DESI:2025zgx}
M.~Abdul Karim \textit{et al.} [DESI],
{\em DESI DR2 Results II: Measurements of Baryon Acoustic Oscillations and Cosmological Constraints},
[arXiv:2503.14738 [astro-ph.CO]].

\bibitem{DESI:2025ejh}
W.~Elbers \textit{et al.} [DESI],
{\em Constraints on Neutrino Physics from DESI DR2 BAO and DR1 Full Shape},
[arXiv:2503.14744 [astro-ph.CO]].

\bibitem{Escudero:2020ped}
M.~Escudero, J.~Lopez-Pavon, N.~Rius and S.~Sandner,
{\em Relaxing Cosmological Neutrino Mass Bounds with Unstable Neutrinos},
JHEP \textbf{12} (2020), 119
[arXiv:2007.04994 [hep-ph]].

\bibitem{Craig:2024tky}
N.~Craig, D.~Green, J.~Meyers and S.~Rajendran,
{\em No \ensuremath{\nu}s is Good News},
JHEP \textbf{09} (2024), 097
[arXiv:2405.00836 [astro-ph.CO]].

\bibitem{Barenboim:2020vrr}
G.~Barenboim, J.~Z.~Chen, S.~Hannestad, I.~M.~Oldengott, T.~Tram and Y.~Y.~Y.~Wong,
{\em Invisible neutrino decay in precision cosmology},
JCAP \textbf{03} (2021), 087
[arXiv:2011.01502 [astro-ph.CO]].

\bibitem{Masso:1999wj}
E.~Masso and R.~Toldra,
{\em Photon spectrum produced by the late decay of a cosmic neutrino background},
Phys. Rev. D \textbf{60} (1999), 083503
[arXiv:astro-ph/9903397 [astro-ph]].

\bibitem{Chianese:2018luo}
M.~Chianese, P.~Di Bari, K.~Farrag and R.~Samanta,
{\em Probing relic neutrino radiative decays with 21 cm cosmology},
Phys. Lett. B \textbf{790} (2019), 64-70
[arXiv:1805.11717 [hep-ph]].

\bibitem{Planck:2018vyg}
N.~Aghanim \textit{et al.} [Planck],
{\em Planck 2018 results. VI. Cosmological parameters},
Astron. Astrophys. \textbf{641} (2020), A6
[erratum: Astron. Astrophys. \textbf{652} (2021), C4]
[arXiv:1807.06209 [astro-ph.CO]].

\bibitem{DiBari:2018vba}
P.~Di Bari,
{\em Cosmology and the Early Universe},
CRC Press, 2018,
ISBN 978-1-4987-6170-3, 978-1-138-49690-3.

\bibitem{Pritchard:2011xb}
J.~R.~Pritchard and A.~Loeb,
{\em 21-cm cosmology},
Rept. Prog. Phys. \textbf{75} (2012), 086901
[arXiv:1109.6012 [astro-ph.CO]].

\bibitem{Zaldarriaga:2003du}
M.~Zaldarriaga, S.~R.~Furlanetto and L.~Hernquist,
{\em 21 Centimeter fluctuations from cosmic gas at high redshifts},
Astrophys. J. \textbf{608} (2004), 622-635
[arXiv:astro-ph/0311514 [astro-ph]].

\bibitem{EDGES}
J.~D.~Bowman, A.~E.~E.~Rogers, R.~A.~Monsalve, T.~J.~Mozdzen and N.~Mahesh,
 {\em An absorption profile centred at 78 megahertz in the sky-averaged spectrum},
  Nature {\bf 555} (2018) no.7694,  67.

\bibitem{Singh:2021mxo}
S.~Singh, J.~Nambissan T., R.~Subrahmanyan, N.~Udaya Shankar, B.~S.~Girish, A.~Raghunathan, 
R.~Somashekar, K.~S.~Srivani and M.~Sathyanarayana Rao,
{\em On the detection of a cosmic dawn signal in the radio background},
Nature Astron. \textbf{6} (2022) no.5, 607-617
[arXiv:2112.06778 [astro-ph.CO]].


\bibitem{2023arXiv230110345B}
S.D.~{Bale}, N.~{Bassett}, J.O.~{Burns}, J.~{Dorigo Jones}, K.~{Goetz},
  C.~{Hellum-Bye} et~al., \emph{{LuSEE 'Night': The Lunar Surface
  Electromagnetics Experiment}},
  \href{https://doi.org/10.48550/arXiv.2301.10345}{\emph{arXiv e-prints} (2023)
  arXiv:2301.10345} [\href{https://arxiv.org/abs/2301.10345}{{\ttfamily
  2301.10345}}].

\bibitem{Fixsen:2009xn}
D.~J.~Fixsen, A.~Kogut, S.~Levin, M.~Limon, P.~Lubin, P.~Mirel, M.~Seiffert, J.~Singal, E.~Wollack and T.~Villela, \textit{et al.}
{\em ARCADE 2 Measurement of the Extra-Galactic Sky Temperature at 3-90 GHz},
Astrophys. J. \textbf{734} (2011), 5
[arXiv:0901.0555 [astro-ph.CO]].

\bibitem{Grebenev:2024nkw}
S.~A.~Grebenev and R.~A.~Sunyaev,
{\em Increase in the Brightness of the Cosmic Radio Background toward Galaxy Clusters},
Astron. Lett. \textbf{50} (2024), 159-185
[arXiv:2408.01858 [astro-ph.HE]].

\bibitem{Dowell:2018mdb}
J.~Dowell and G.~B.~Taylor,
{\em The Radio Background Below 100 MHz},
Astrophys. J. Lett. \textbf{858} (2018) no.1, L9
[arXiv:1804.08581 [astro-ph.CO]].

\end{thebibliography}
\end{document}